\documentclass[superscriptaddress,amsmath,amssymb,aps,pra]{revtex4-2}

\usepackage{graphicx}
\usepackage{dcolumn}
\usepackage{bm}
\usepackage[colorlinks]{hyperref}

\begin{document}

\title{The identification of mean quantum potential with Fisher information leads to a strong uncertainty relation}

\author{Yakov Bloch}
\affiliation{Faculty of Engineering and the Institute of Nanotechnology and Advanced Materials, Bar-Ilan University, Ramat Gan 5290002, Israel}

\author{Eliahu Cohen}
\affiliation{Faculty of Engineering and the Institute of Nanotechnology and Advanced Materials, Bar-Ilan University, Ramat Gan 5290002, Israel}

\begin{abstract}
The Cram\'er-Rao bound, satisfied by classical Fisher information, a key quantity in information theory, has been shown in different contexts to give rise to the Heisenberg uncertainty principle of quantum mechanics. In this paper, we show that the identification of the mean quantum potential, an important notion in Bohmian mechanics, with the Fisher information, leads, through the Cram\'er-Rao bound, to an uncertainty principle which is stronger, in general, than both Heisenberg and Robertson-Schr\"odinger uncertainty relations, allowing to experimentally test the validity of such an identification.
\end{abstract}

\maketitle

\section{Introduction}
The paper analyzes and utilizes the relation between the mean quantum potential appearing in de Broglie-Bohm theory and the Fisher information. It could therefore be helpful to first review these two concepts.

\subsection{de Broglie-Bohm Theory}

The de Broglie-Bohm formulation of quantum theory \cite {bohm1952I, bohm1952II, undivided, Holland, Durr2009, Cushing2013} is a realistic and deterministic framework for the description of quantum phenomena, allowing the description of individual quantum events \cite{Howtoteachquantum}. The theory is sometimes called ontological, since it attempts to speak about what exists, rather than what one can measure \cite{ont}. It does so by the introduction of (nonlocal) ``hidden variables''- properties of the quantum particle that cannot be measured, but are, rather, asserted. Such are the positions of the Bohmian particles in the de Broglie-Bohm theory. Unlike standard quantum theory, which attributes the statistical properties of the ensemble to the individual particle, giving up the concreteness of position and momentum \cite{sep-qm-copenhagen}, in the de Broglie-Bohm theory, the latter is retained while the former is forsaken- the particle’s behavior is not inherently probabilistic \cite{Bohmchance}, but, rather, essentially deterministic. In Bohm's theory, the probabilistic features express a lack of knowledge about the initial conditions. An ensemble of such particles would reproduce the standard quantum distribution, given by the Born rule, 
\begin{equation}
    \rho = |\psi|^2
    \label{eq:Born}
\end{equation}
Having a specific position at all times, independent of the measurement process, the particle’s momentum is assumed to be proportional to the local wavenumber \cite{Berrylocalwave},
\begin{equation}
    \vec{k} = \text{Im} \{\nabla \text{ln}\psi\}
    \label{eq:localwavenumber}
\end{equation}
which constitutes the Bohmian guiding equation \cite{sep-qm-bohm},
\begin{equation}
    \frac{d\vec{x}}{dt} = \frac{\hbar}{m} \text{Im} \{\nabla \text{ln}\psi\}
    \label{eq:guidance}
\end{equation}
As the local wavenumber is derived from the wavefunction, the particle is said to be guided by the wave. The guiding equation does not depend on the specific equation satisfied by the wavefunction, according to which it evolves. Having defined the particle’s momentum as such, the Schr\"odinger equation,
\begin{equation}
    i\hbar\  \frac{\partial\Psi}{\partial t}=\ -\frac{\hbar^2}{2m}\ \mathrm{\nabla}^2\ \mathrm{\Psi}+V\mathrm{\Psi}
    \label{eq:Schrodinger}
\end{equation}
becomes an equation describing the guiding process. A polar decomposition of the equation (after Madelung) yields a probability conservation equation for the distribution of particles and a modified Hamilton-Jacobi equation written in terms of the local or Bohmian momentum cite{Madelung}. This is a simple manipulation of the equation that consists of inserting a polar decomposition of the wavefunction  
\begin{equation}
    \Psi= Re^{i \frac{S}{\hbar}}
    \label{eq:polar}
\end{equation}
The two equations are the continuity equation,
\begin{equation}
   \frac{\partial(R^2)}{\partial t}+\nabla\left(R^2\frac{\nabla S}{m}\right)=0
    \label{eq:cont}
\end{equation}
And the modified Hamilton-Jacobi equation,
\begin{equation}
   \frac{\mathrm{\partial S}}{\mathrm{\partial t}}+\frac{\left(\nabla S\right)^2}{2m}+V-\frac{\hbar^2}{2m}\frac{\nabla^2R}{R}=0
    \label{eq:HJ}
\end{equation}
Where, according to (3), the local, or, Bohmian, momentum is simply 
\begin{equation}
   \vec{P_q} = \nabla S
    \label{eq:localmomentum}
\end{equation}
This equation is called the quantum Hamilton-Jacobi equation, and it differs from its classical counterpart by an extra term called the quantum potential \cite{Quantpot}.
\begin{equation}
   Q=\ -\frac{\hbar^2}{2m}\frac{\nabla^2R}{R}
    \label{eq:quantpot}
\end{equation}
This term is interpreted, in the de Broglie-Bohm theory, as an extra potential, from which a force that acts on the particles, changing their velocities and bending their trajectories, is derived. This quantum force mediates the influence of the wave on the particle, describing the guidance mechanism. This extra term, the quantum potential, accounts for all quantum phenomena, and is highly nonlocal, enfolding information about the whole experimental setup. The quantum potential involves a multiplicative constant, $\frac{\hbar^2}{2m}$, such that, at the limit of $\hbar \rightarrow 0$ , the classical Hamilton-Jacobi equation is retrieved. In this limit, quantum effects, which in the Bohmian perspective are interpreted as the influence of the wave on the particle, become negligible, and the dynamics is classical.   

\subsection{Fisher information}

Fisher information \cite{Fisher1, Fisher2, fisher3} is a fundamental quantity in the theory of information, which measures the amount of information that an observable random variable $Y$ carries about an unknown parameter $\theta$ upon which the probability of $Y$ depends. The likelihood $\rho(y|\theta)$ is the probability density function for $y$ conditioned on the value of $\theta$. In terms of the likelihood, the Fisher information is given by:
\begin{equation}
   I(\theta) \equiv \int \frac{1}{\rho(y|\theta)}\left(\frac{\partial \rho(y|\theta)}{\partial\theta}\right)^2 dy.
    \label{eq:Fisher}
\end{equation}
The quantum version of Fisher information \cite{Helstrom, quantumfisher} has been extensively used in quantum metrology \cite{metro} and statistical inference \cite{inference}, and has been used in the context of entanglement detection \cite{Entangle}.
The random variable $\hat{Y}$ is an unbiased estimator for the parameter $\theta$ if
\begin{equation}
   E_{\theta}[\hat{Y}] = \theta.
    \label{eq:unbiased}
\end{equation}
That is, the expectation value of the estimator is the parameter. The variance of an unbiased estimator is bounded from below by the inverse of fisher information, in what is known as the Cram\'er-Rao bound \cite{kay1993fundamentals}:
\begin{equation}
   \text{Var}[\hat{Y}] \geq 1/I(\theta).
    \label{eq:CR}
\end{equation}
We show that this bound leads, through the connection with the quantum potential, to an uncertainty relation, generally stronger than the Robertson-Schr\"odinger uncertainty relation \cite{Robertson}.

\section{The mean quantum potential and Fisher information}

The mean quantum potential of Bohmian mechanics has been related \cite{Derivation, Quantproper, Towarda, OnEntropy, frieden}, on the basis of formal similarity, to Fisher information, a quantity central in information theory, by the following formula,
\begin{equation}
    \bar{Q} = \frac{\hbar^2}{8m} I,
    \label{eq:polarexpreal1}
\end{equation}
where $\bar{Q}$ is the mean quantum potential and I is the Fisher information about the observable $\hat{x}$. This connection has been often asserted due to formal similarity of the definitions \cite{Towarda}. However, since the two concepts arise in different contexts and theories, albeit both closely related to measurement, this connection has to be physically justified. Even more so, noticing that the definition of Fisher information includes conditional probabilities, absent from the expression of the mean quantum potential. In other words, to justify this connection, one has to reduce the former to the latter, providing concrete justification for such a reduction. An attempt for such a justification is outlined in Reginatto’s paper \cite{Derivation} serving as the background for the derivation of the Schr\"odinger equation from a variational principle of minimum Fisher information. He describes the measurement problem of quantum mechanics as an estimation problem of a deterministic variable with a superimposed random noise. That is, estimating the parameter $\theta$ in the presence of unknown added noise $x$, a measurement $y$ of the parameter is related to $x$ and $\theta$ by
\begin{equation}
   y = \theta + x,
    \label{eq:Noise}
\end{equation}
where $y$ is a measurement of the particle's position, while $\theta$ is its actual ``hidden" position.
With the addition of an assumption regarding the conditional probability distribution of the measured outcome given the actual value, the assumption of translation invariance, formally described as,
\begin{equation}
   \rho(y|\theta) = \rho(y-\theta) = \rho(x).
    \label{eq:Density}
\end{equation}
And so, Fisher information becomes,
\begin{equation}
   I(\theta) = \int \frac{1}{\rho(x)}\left(\frac{\partial \rho(x)}{\partial\theta}\right)^2 dx = -\int \rho \frac{\partial^2 \text{ln} \rho(x)}{\partial \theta^2} dx.
    \label{eq:FisherX}
\end{equation} 
Using integration by parts one can show,
\begin{equation}
   \int \rho \frac{\partial^2 \text{ln} \rho(x)}{\partial x^2} dx = 4 \int \rho \frac{1}{R} \frac{\partial^2 R}{\partial x^2} dx = -\frac{8m}{\hbar^2}\bar{Q},
    \label{eq:meanQP}
\end{equation} 
so that, if one calculates $I(x)$, that is, Fisher information about the position, one obtains equation (\ref{eq:polarexpreal1}), relating the mean quantum potential with Fisher information.
Let us first notice that what Reginatto describes is a theory of measurement for hidden variables, recasting the measurement problem into an estimation problem in which the concept of Fisher information naturally arises. What is the nature of the noise in the measurement and its source? This question is not addressed, and yet, translation invariance is assumed about the conditional probability distribution. What is the justification of such an assumption? A possible justification, outlined by Frieden \cite{Frieden2} is the fact that the measurement does not depend on the position in which it is performed. One might perform the experiment in a different position without expecting different results. However, it is clear that for the identification of Fisher information with the quantum potential, one must have,
\begin{equation}
   \left(\frac{\partial \rho(x)}{\partial \theta}\right)^2 = \left(\frac{\partial \rho(x)}{\partial x}\right)^2.
    \label{eq:condforquant}
\end{equation}
While the previous assumptions are partially justified in the literature, this assumption goes unnoticed. This statement is equivalent to,
\begin{equation}
   \left(\frac{\partial x(\theta)}{\partial \theta}\right)^2 = 1,
    \label{eq:condforquant}
\end{equation}
which means that an increment of the ``hidden" position results in an equal increment in the noise. In other words, the noise is proportional to the actual position of the particle. What is the justification for such an assumption? It is not at all certain that this assumption can be justified. In this paper we propose an experimental test for all of these assumptions. Taking the connection with Fisher information seriously, we use the Cram\'er-Rao bound to arrive at an uncertainty principle, stronger than the Robertson-Schr\"odinger. Breaking this uncertainty would imply that it is inappropriate to call the mean quantum potential ``Fisher information”, which means that one or more of Reginatto's assumptions are false. It is important to note that Reginatto’s reasoning in the rest of the paper is not hurt by our criticism, as Fisher information does not lend the derivation any of its properties other than its name. The quantity minimized would simply be renamed. It is our goal simply to check whether the name, which, again, should connect two distinct theories in a nontrivial way, is appropriate. Had it not been falsified, the study would not only provide a tighter bound on uncertainty through information theory, but also suggest and motivate further research in this quantum triple point, connecting quantum measurement theory, estimation theory and nonlocal hidden variables, justifying experimentally Reginatto's assumptions about quantum measurement and nonlocal hidden variables. The rest of the paper is organized as follows:        

\section{Connecting Bohmian quantities to quantum observables}

The expectation value of momentum is given by the expression
\begin{equation}
     \langle \hat{P} \rangle = \langle\psi|\hat{P}|\psi\rangle.
    \label{eq:expectation}
\end{equation}
To write this expression in the polar representation, let us insert the resolution of the identity
\begin{equation}
     \langle \hat{P} \rangle = \int \langle\psi|\vec{r}\rangle\left\langle \vec{r}\right|\hat{P}\left|\psi\right\rangle d^3r.
    \label{eq:expectationwithidentity}
\end{equation}
The polar representation of the wavefunction is 
\begin{equation}
    \langle \vec{r}|\psi\rangle = Re^{i\frac{S}{\hbar}},
    \label{eq:polar}
\end{equation}
with $\hat{P} = -i\hbar\nabla$ we have
\begin{equation}
    \langle \hat{P} \rangle = \int R^2\left(-i\hbar\frac{\nabla R}{R} + \nabla S\right) d^3r.
    \label{eq:polarexp}
\end{equation}
This expression has two parts: an expectation value of the osmotic momentum and of the Bohmian momentum. The first of the two can be shown to be equal to zero when the expectation is evaluated over the whole range, and the wavefunction tends to zero at the boundaries. This is a consequence of the gradient Gauss theorem:
\begin{equation}
    \int_V -i\hbar R\nabla R d^3r = \int_V -i\hbar \frac{1}{2} \nabla \rho d^3r = \int_{\partial V} -i\hbar \frac{1}{2} \rho \vec{dS} = 0.
    \label{eq:gradientGauss}
\end{equation}
And so, we have that the expectation value of the momentum operator is equal to the mean value of the Bohmian, or local, momentum, that is
\begin{equation}
    \langle \hat{P} \rangle = \bar{P_q}.
    \label{eq:equalityofexpectations}
\end{equation}
Following the same procedure as with the expectation of momentum, we get, 
\begin{equation}
    \langle \hat{P^2} \rangle = \bar{P_q^2} + 2m\bar{Q},
    \label{eq:polarexpreal}
\end{equation}
where $Q$ is the quantum potential.
In \cite{} the expectation value of the quantum potential is related to the Fisher information, $I$, about the observable $\hat{x}$, by, 
\begin{equation}
    \bar{Q} = \frac{\hbar^2}{8m} I.
    \label{eq:polarexpreal2}
\end{equation}
And so, what we have is a relation between the expectation value of the momentum operator squared, the mean squared local (Bohmian) momentum and Fisher information. Let us rewrite it in terms of I:
\begin{equation}
    \langle \hat{P^2} \rangle = \bar{P_q^2} + \frac{\hbar^2}{4} I.
    \label{eq:Fisher}
\end{equation}
Using the previous results, we can write the variance of the momentum operator as 
\begin{equation}
    \text{Var}(\hat{P}) = \langle \hat{P^2} \rangle - \langle \hat{P} \rangle^2 = \bar{P_q^2} - \bar{P_q}^2 + \frac{\hbar^2}{4} I.
    \label{eq:Var}
\end{equation}

That is, 
\begin{equation}
    \text{Var}(\hat{P}) = \text{Var}(P_q) + \frac{\hbar^2}{4} I,
    \label{eq:Var}
\end{equation}
which means that the difference between the variance of the momentum operator and the local momentum is proportional to the Fisher information.
Using the Cram\'er-Rao bound and the relationship between the variance of the momentum operator, the variance of the local momentum and Fisher information, we find a tighter bound on uncertainty, of the form
\begin{equation}
    \text{Var}(\hat{P})\text{Var}(\hat{X}) \geq \frac{\hbar^2}{4} + \text{Var}(P_q)\text{Var}(\hat{X}),
    \label{eq:BetterHeisenberg}
\end{equation}
which, for zero local momentum variance, becomes identical to the Heisenberg uncertainty.
\begin{equation}
    \text{Var}(\hat{P})\text{Var}(\hat{X}) \geq \frac{\hbar^2}{4}.
    \label{eq:BetterHeisenberg}
\end{equation}

In fact, this result, namely, the derivation of the Heisenberg uncertainty principle, has been achieved by Reginatto himself \cite{reginatto2}. Others, as early as A. J. Stam in 1959, have derived the Heisenberg uncertainty principle from the Cram\'er-Rao bound through different considerations \cite{Stam, FisherHeis, Fisherheis2, Fisherheis3}. In \cite{Tighter}, tighter uncertainty relations were formulated for mixed states by replacing one variance by the quantum Fisher information. Our approach, namely, that of connecting Bohmian quantities to quantum observables and following the consequences of the relation of Fisher information to mean quantum potential, leads to an uncertainty principle tighter than Robertson-Schr\"odinger, without any modification of known uncertainty relations. To see that, let us rewrite the Robertson-Schr\"odinger uncertainty principle in Bohmian terms. In its general form, the Robertson-Schr\"odinger uncertainty principle is given by \cite{Tighter, Robertson},
\begin{equation}
    \text{Var}_{\rho}(A)\text{Var}_{\rho}(B) \geq \frac{1}{4}|Tr(\rho[A,B])|^2 + |\text{Re}\{\text{Cov}_{\rho}(A,B)\}|^2.
    \label{eq:RS}
\end{equation}
Now, since,
\begin{equation}
    \text{Re}\{\text{Cov}_{\rho}(\hat{P},\hat{X})\} = \frac{1}{2}\langle \{\hat{P},\hat{X}\} \rangle -\langle \hat{P} \rangle \langle \hat{X} \rangle,
    \label{eq:Recov}
\end{equation}
the anti-commutator of momentum and position can be written as
\begin{equation}
    \{\hat{P},\hat{X}\} = -i\hbar + 2\hat{X}\hat{P}.
    \label{eq:anticom}
\end{equation}
And so, we can write
\begin{equation}
    \text{Re}\{\text{Cov}_{\rho}(\hat{P},\hat{X})\} = -i\frac{\hbar}{2} + \langle \hat{X}\hat{P} \rangle - \langle \hat{P} \rangle \langle \hat{X} \rangle.
    \label{eq:Recovexpectations}
\end{equation}
Writing the integral explicitly and using the polar decomposition of the wavefunction, the expectation value of the product of position and momentum becomes,
\begin{equation}
    \langle \hat{X}\hat{P} \rangle = \int R^2 x \left(-i\hbar \frac{1}{R}\frac{\partial R}{\partial x} + \frac{\partial S}{\partial x}\right)dx.  
    \label{eq:xpexp}
\end{equation}
Integrating by parts and using the fact that,
\begin{equation}
    \langle \hat{P} \rangle = \bar{P_q}.
    \label{eq:equalityofexpectations}
\end{equation}
We finally arrive at
\begin{equation}
    \text{Re}\{\text{Cov}_{\rho}(\hat{P},\hat{X})\} = \text{Cov}(x,p_q).
    \label{eq:Recovexpectations}
\end{equation}
And so, the Robertson-Schr\"odinger uncertainty can be written as,
\begin{equation}
    \text{Var}(\hat{X})\text{Var}(\hat{P}) \geq \frac{\hbar^2}{4} + \text{Cov}^2(x,P_q)
    \label{eq:RSU}
\end{equation}
(coinciding with the Heisenberg uncertainty relation when $x$ and $P_q$ are uncorrelated).

\section{Relation to quantum uncertainties}
For zero local momentum variance, the Cram\'er-Rao uncertainty becomes identical to the Heisenberg uncertainty. Since the Cram\'er-Rao bound assumes its minimum for a normal distribution and so does the Heisenberg uncertainty, which is minimal for a Gaussian wavepacket, for which the variance of local momentum is zero, this result is very intuitive. More generally, we can say that the two are equivalent whenever the local momentum has zero variance.
Written in Bohmian terms, the Robertson-Schr\"odinger uncertainty principle assumes the form given by Eq. \ref{eq:RSU}. Thus, the Robertson-Schr\"odinger uncertainty relation is related to the one that follows from the Cram\'er-Rao bound by the Cauchy-Schwartz inequality,
\begin{equation}
    \text{Cov}^2(x,P_q) \leq \text{Var}(x)\text{Var}(P_q),
    \label{eq:CS}
\end{equation}
such that the former is always weaker or at most equivalent to the latter, with the inequality saturating when the Pearson correlation coefficient of $x$ and $P_q$ is equal to $\pm 1$. Intuitively, we may understand this result as an epistemic indifference to the position dependence of Bohmian momentum, coming from information theory. This new bound leaves a range of uncertainties forbidden according to the combination of quantum theory and estimation theory. This range, whose width is
\begin{equation}
    \Delta = \text{Var}(x)\text{Var}(P_q) - \text{Cov}^2(x,P_q)  
    \label{eq:CSdiff}
\end{equation}
could allow to test whether it is appropriate to treat the quantum measurement process as a classical estimation task.

\section{Conclusions}

There has been a growing literature mentioning a connection between the quantum potential and Fisher information. Physical justification for the assumptions required for this identification is usually absent, and the relation is often treated simply as a mathematical identity although it might be more fundamental. To us it seems that a statement regarding a direct, profound connection between two such remote quantities originating from two distinct theories (even if both are eventually concerned with the process of measurement) ought to be a strong one, and hence has to be carefully examined.

Using the tools of information theory, to which the quantum potential should be supposedly connected, we devised an experimental test which should allow to falsify this connection. As the inverse Fisher information is the lower bound of the variance of an unbiased estimator, known as the Cram\'er-Rao bound, writing the variance of momentum using the mean quantum potential, and assuming its relation to Fisher information, we arrive at an uncertainty relation, that is, to an inequality bounding from below the product of the variances of position and momentum, which is in general stronger than the Robertson-Schr\"odinger uncertainty principle. Equivalence between the two is established when the  Pearson correlation coefficient of the position and the Bohmian momentum is equal to $\pm 1$ (and equivalence to the Heisenberg uncertainty relation is achieved whenever the position and Bohmian momentum are uncorrelated). The uncertainty relation we derived provides a range of uncertainties forbidden by information theoretic constraints applied to quantum mechanics. Measuring an uncertainty within this ``forbidden'' region would suggest that equating the mean quantum potential with Fisher information is inappropriate. On the contrary, had the relation not been falsified, our study would suggest, in general, a tighter bound on uncertainty, and some reason to believe that there could indeed exist a relation between quantum measurements, classical estimation theory and nonlocal hidden variables (in the form used by Reginatto). 

In future work it could be of interest to generalize the proposed uncertainty relation to other physical variables and moreover to exploit the affinity between uncertainty and nonlocal correlations \cite{Carmi2019} for deriving similar bounds on quantum correlations exhibited by entangled states.

\section*{Acknowledgements}

This research was supported by Grant No. FQXi-RFP-CPW-2006 from the Foundational Questions Institute and Fetzer Franklin Fund, a donor-advised fund of Silicon Valley Community Foundation. E.C. was supported by the Israeli Innovation Authority under Projects No. 70002 and No. 73795, by the Pazy Foundation, by ELTA Systems LTD - Israel Aerospace Industries (IAI) division, by the Israeli Ministry of Science and Technology, and by the Quantum Science and Technology Program of the Israeli Council of Higher Education.

\bibliography{citations}

\end{document}